\documentclass[5p,sort&compress]{elsarticle}
\usepackage{amssymb,amsmath,amsfonts,psfrag}
\usepackage{graphicx}                                 
\newcommand{\Tr}{\operatorname{Tr}}
\renewcommand{\Re}{\mathfrak{Re}\,}
\newcommand{\Fig}[1]{Fig.~\ref{#1}}

\newcommand{\Sec}[1]{Sec.~\ref{#1}}
\newcommand{\Eq}[1]{Eq.~(\ref{#1})}
\graphicspath{{./Figures/}}
\newcommand{\preprint}{\setlength{\unitlength}{1mm}{\begin{picture}(0,0)
  \put(132,15){\mbox{\footnotesize%
      HU-EP-08/69, ADP-08-18/T678}
}\end{picture}}}

\begin{document}

\title{\preprint 
Lattice gluodynamics computation of Landau-gauge Green's
  functions\newline in the deep infrared}

\author[moscow]{I.L.~Bogolubsky}
%
\author[berlin]{E.-M.~Ilgenfritz}
%
\author[berlin]{M.~M\"uller--Preussker}
%
\author[adelaide]{A.~Sternbeck}

\address[moscow]{Joint Institute for Nuclear Research, 141980 Dubna,
  Russia}
\address[berlin]{Humboldt-Universit\"at zu Berlin, Institut f\"ur Physik,
  12489 Berlin, Germany}
\address[adelaide]{CSSM, School of Chemistry \& Physics, University of
  Adelaide, SA 5005, Australia}

\date{May 14, 2009}

\begin{abstract}
We present recent results for the Landau-gauge gluon and ghost propagators 
in $SU(3)$ lattice gluodynamics obtained on a sequence of
lattices with linear extension ranging from $L=64$ to $L=96$ at 
$\beta = 5.70$, thus reaching ``deep infrared'' momenta down to $75$ MeV. 
Our gauge-fixing procedure essentially uses a simulated annealing 
technique which allows us to reach gauge-functional values closer to the 
global maxima than standard approaches do. Our results are  
consistent with the so-called decoupling solutions found for 
Dyson-Schwinger and functional renormalization group equations.
\end{abstract}


\maketitle

\section{Introduction}
\label{intro}

The infrared behaviour of gauge-variant Green's functions of Yang-Mills
theories has become an increasingly interesting topic during the last
decade. The interest was originally stimulated by confinement
scenarios proposed a long time ago by Kugo and
Ojima~\cite{Ojima:1978hy,Kugo:1979gm}, Gribov~\cite{Gribov:1977wm} and
Zwanziger~\cite{Zwanziger:1991gz,Zwanziger:1993dh}. The recent
progress is due to the discovery of consistent asymptotic solutions of
the whole tower of Dyson-Schwinger (DS) equations and more recently of
functional renormalization group (FRG) equations in the deep infrared
(IR) limit~\cite{vonSmekal:1997is,vonSmekal:1997vx,Lerche:2002ep,
Zwanziger:2001kw,Pawlowski:2003hq,Fischer:2006vf,Fischer:2008uz}. 
These solutions called \emph{scaling} or \emph{conformal} solutions 
behave according to power laws with well-determined 
exponents~\cite{Lerche:2002ep,Zwanziger:2001kw,Fischer:2006vf}. The
expectation is that they respect global BRST invariance. Since these
solutions lead to a vanishing gluon propagator and correspondingly to
an IR-singular ghost dressing function they fit nicely with the
aforementioned scenarios. The running coupling related to the
ghost-ghost-gluon vertex exhibits an IR fixed
point~\cite{vonSmekal:1997is} as also proposed by Shirkov and
Solovtsov \cite{Shirkov:1997wi,Shirkov:2002gw}.

There is a different set of so-called \emph{decoupling} solutions as
proposed in
\cite{Boucaud:2007va,Dudal:2007cw,Aguilar:2008xm,Boucaud:2008ji,Boucaud:2008ky}.
These solutions --- thoroughly discussed also in \cite{Fischer:2008uz}
--- are characterised by a non-zero IR gluon propagator as well as by
an IR-finite ghost dressing function, i.e., they do not agree with the
Kugo-Ojima criterion.  The name \emph{decoupling} reflects the fact
that the corresponding running coupling decreases towards zero in the
limit of vanishing momenta. But this behaviour does not mean that the
\emph{decoupling solutions} contradict gluon and quark confinement
(see also the discussion in \cite{Fischer:2008yv}).  Both sets of
solutions demonstrate the expected positivity violation of the gluon
propagator as well as providing the expected Polyakov loop behaviour at
the deconfinement transition in pure $SU(2)$ and $SU(3)$ gauge
theories~\cite{Braun:2007bx}. Moreover, both types of solutions, when
interpolated from the infrared asymptotics to the perturbative region
by solving numerically the (properly truncated) system of DS or FRG
equations, behave quite similarly in the momentum range relevant for
hadron phenomenology.

Even if it might appear to be an academic question, it is important to
ask which set of solutions is the correct one. Ab-initio lattice gauge
theory computations are expected to solve this issue. $SU(2)$ and
$SU(3)$ lattice computations of Landau-gauge gluon and ghost
propagators have been carried out by several groups. For a recent
review see \cite{Cucchieri:2007md} and papers cited therein. On the
one hand, as long as in the four-dimensional case the linear lattice
sizes did not reach far beyond $O(5~\mathrm{fm})$ the ghost dressing
function was observed to rise towards the infrared limit. When fitted
with a power law the corresponding exponent was found to be much
smaller than predicted by the scaling solutions. On the other hand, the
gluon propagator has not yet been found to decrease towards the IR limit
but rather approaching a finite plateau value at $p=0$ (for $SU(3)$
see~\cite{Sternbeck:2005tk,Sternbeck:2006cg,Ilgenfritz:2006he}). However,
investigations of the DS equations on a 4D torus \cite{Fischer:2007pf}
have demonstrated that linear box sizes of $O(10~\mathrm{fm})$ or even
more, might be necessary in order to see the correct asymptotic
behaviour.

In the meantime simulations of $SU(2)$ pure lattice gauge theory have
reached huge lattice sizes \cite{Cucchieri:2007md,Sternbeck:2007ug}
confirming nothing but an IR-plateau behaviour for the gluon
propagator. Moreover, in \cite{Cucchieri:2007rg} Cucchieri and Mendes
derived infrared bounds for the gluon and ghost propagators
indicating, e.g., a non-zero value for the zero-momentum gluon
propagator in the infinite-volume extrapolation.

However, the aforementioned results of DS and FRG studies as well as most of 
the lattice results obtained with the standard \emph{minimal} Landau gauge 
did not take into account the effect of Gribov copies. 
For $SU(2)$ some of us have shown \cite{Bakeev:2003rr,Bogolubsky:2005wf,
Bogolubsky:2007pq,Bogolubsky:2007bw,Bornyakov:2008yx} that the influence of 
Gribov copies is more severe than many authors might have expected
(see also the recent investigation in \cite{Maas:2008ri}).
The effect is even stronger if non-periodic $Z(2)$ gauge-transformations 
(\emph{flips} of all link variables orthogonal to fixed 3D sheets with a 
factor -1) are taken into account. Already on modest lattice volumes the 
gluon propagator as well as the ghost dressing function were seen to run 
into IR plateaus indicating that lattice results seem to support the 
\emph{decoupling solution} of DS and FRG equations \cite{Bornyakov:2008yx}.

In the $SU(3)$ case the effect of Gribov copies on the Landau gauge gluon 
and ghost propagators has been studied without $Z(3)$ flips so far
\cite{Silva:2004bv,Sternbeck:2005tk,Sternbeck:2005vs} showing that
only the ghost propagator seems to be systematically affected within 5
to 10\,\%.

In the meantime, we have extended the computation of Landau gauge gluon 
and ghost propagators in $SU(3)$ gluodynamics to linear lattice sizes of
around $O(16~\mathrm{fm})$. 
Preliminary results for lattice sizes up to $80^4$ have been presented 
already in \cite{Bogolubsky:2007ud} clearly showing that an infrared 
plateau of the gluon propagator evolves and demonstrating --- to our
knowledge for the first time --- a flattening of the ghost
dressing function as well. Here we go a step further in order to get more
confidence by increasing the lattice size for the gluon propagator
up to as much as $96^4$ as well as by increasing considerably our statistics
for both propagators. In our investigation we have
put some emphasis on careful gauge fixing with the \emph{simulated 
annealing} method which wins in efficiency in comparison with standard 
over-relaxation, the more degrees of freedom the system has (see below and 
discussions in \cite{Bogolubsky:2007pq}).

\section{General setup}
\label{setup}

We compute the $SU(3)$ gluon and ghost propagators with Monte Carlo
(MC) techniques on a lattice with periodic boundary conditions. The
standard Wilson single-plaquette action and the lattice definition for
the gauge potentials
\begin{equation}
 A_\mu(x+\hat{\mu}/2)
  :=\frac{1}{2iag_0}\left(U_{x\mu} -
  U^{\dagger}_{x\mu}\right)\Big|_{\mathrm{traceless}}
\label{eq:gaugepotential}
\end{equation}
are adopted. In order to fix the Landau gauge for each lattice gauge field 
$\{U\}$ generated by means of the MC procedure, the gauge functional
\begin{equation}
  F_{U}[g] = \frac{1}{3}\sum_{x}\sum_{\mu=1}^{4}\Re\Tr\; g_x U_{x\mu}
  \,g^{\dagger}_{x+\mu} 
\label{eq:gaugefunctional}
\end{equation}
is iteratively maximised with respect to a gauge transformation
$\{g_x\}$ which is taken as a periodic field as well.
In order to approach the global maximum (related to the
fundamental modular region) as closely as possible, we use the
\emph{simulated annealing} (SA) algorithm~\cite{Kirkpatrick:1983aa}, in
combination with subsequent standard over-relaxation (OR). The
latter is applied in the final stage of the gauge-fixing procedure
in order to finalise the transformation to any required precision
of the transversality condition $\nabla_{\mu} A_{\mu} = 0$. More
than a decade ago, SA was shown to be very efficient when dealing
with the maximally Abelian gauge~\cite{Bali:1994jg,Bali:1996dm}.
\begin{figure}[t]
  \centering
  \includegraphics[width=0.95\linewidth]{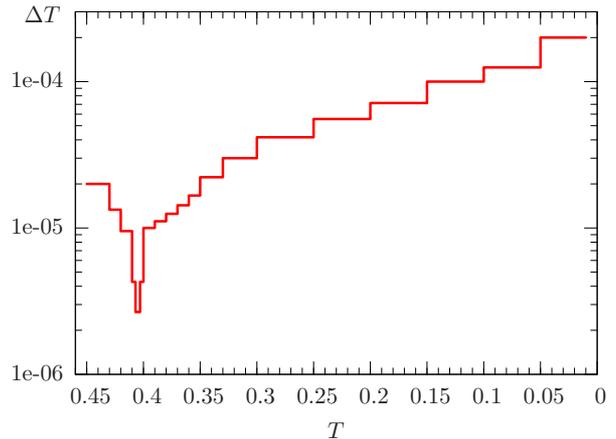}
  \caption{Example of a SA schedule. Shown is the
    temperature step size $\Delta T$ by which $T$ is
    reduced in the SA algorithm used on a $96^4$~lattice (see
    text for further details).}
  \label{fig:sa_schedule96}
\end{figure}

In case of the present approach the SA algorithm  generates a
field of gauge transformations $g_x$ by MC iterations with a
statistical weight proportional to $\exp(F_U[g]/T)$. The
\emph{temperature} $T$ is an auxiliary parameter which is gradually
decreased in order to maximise the gauge functional $F_U[g]$.
In the beginning, the $T$ chosen must be large enough 
to allow the traversing of the configuration space of $g_x$
fields in large steps. An initial value $T_{\rm init}=0.45$ 
was found to be sufficient for that~\cite{Schemel:2006xx}. After 
each sweep, including one heatbath and four micro-canonical update 
steps at each lattice site, $T$ is decreased
until $g_x$ is captured in a particular basin of attraction. 
We choose the lowest temperature value to be $T_{\mathrm{final}}=0.01$
and use a fine-tuned simulated annealing schedule before applying 
final over-relaxation steps to reach Landau gauge with good precision 
($\max_x \Re\Tr[(\nabla_{\mu}A_{x\mu})(\nabla_{\mu}
A^{\dagger}_{x\mu})] < 10^{-13}$).

An infinitely slow and long simulated annealing process would
definitely lead to the global maximum. In practice, however, we chose
schedules according to which the temperature is reduced in
steps of finite but varying size. These schedules are motivated by the
following observation. When the thermal average $\langle F_U[g]
\rangle$ is considered as a 
function of $T$, one easily notices its monotonous but non-linear
dependence.  In fact, $\langle F_U[g] \rangle$ increases when $T$
decreases, but its gradient reaches a strong maximum at a ``critical''
value $T \simeq 0.405 \pm 0.01$, which resembles very much a phase
transition~\cite{Schemel:2006xx}. This means that after starting with
small $\Delta T$ steps, as required for proper thermalisation close to
$T_{\rm init}=0.45$, it is important to keep the step size especially
small within the narrow critical region $T \simeq 0.40 \ldots
0.41$. Indeed, the distribution of the \emph{final} gauge-functional
values (those of the local maxima after simulated annealing and
over-relaxation) is shifted to noticeably higher values if the
temperature step size $\Delta T$ is reduced according to the rise of
$\langle F_U[g] \rangle$, i.e., if $\Delta T$ is taken relatively
small within the critical interval. This is in particular an efficient 
approach, as almost no improvement was observed, when further reducing
$\Delta T$ outside this interval~\cite{Schemel:2006xx}. 

As an example, \Fig{fig:sa_schedule96} shows the step sizes we use
for the different temperature intervals on a $96^4$ lattice, where,
especially around $T\simeq0.40\ldots0.41$, $\Delta T$ is taken to be very
small. Such extremely long SA 
``chains'', with an order $O(10^4)$ of iterations, allow us with high
probability to reach local   
maxima of the gauge functional $F_U[g]$ close to the \emph{global}
maximum (as far as possible for the computer resources available),
i.e., to the fundamental modular region, even with only one
gauge-fixing attempt (``first-copy approach'').

Note that we do not apply here $Z(3)$ flips, the so-called FSA 
method~\cite{Bogolubsky:2005wf,Bogolubsky:2007bw} which in principle 
is capable of providing even larger $F_U[g]$ values than
our SA algorithm does. But we have seen the difference between SA 
and FSA results decreases when $L$ increases~\cite{Bogolubsky:2007bw}.

The computations presented are carried out at rather strong coupling,
$\beta = 6/g_0^2 = 5.70$. The reason for this choice is to get access
to comparatively large physical volumes at the price of a rather
coarse lattice ($a\approx0.17\,\mathrm{fm}$). In order to study the
volume dependence we 
calculate the gluon propagator for linear lattice sizes $L$ ranging
from $64$ to $96$. Thus, our largest lattice size corresponds to
$(16~\mathrm{fm})^4$. The ghost propagator is studied for linear
lattice sizes $L=64$ and $L=80$. At larger lattice sizes
and the lowest momenta the inversion of the Faddeev-Popov matrix
turns out to converge only in rare cases. Obviously, this is due to
the occurrence of very small eigenvalues which generate some
algorithmic problems. There, a modification of the used matrix
inversion method (see \Sec{ghost}) or an even better gauge fixing,
driving configurations further away from the Gribov horizon
\cite{Sternbeck:2005vs}, could be valuable means of reducing such
problems in future, but it is beyond the scope of the current work.

\section{Gluon propagator}
\label{gluon}

\begin{figure}
  \centering
  \includegraphics[height=6.7cm]{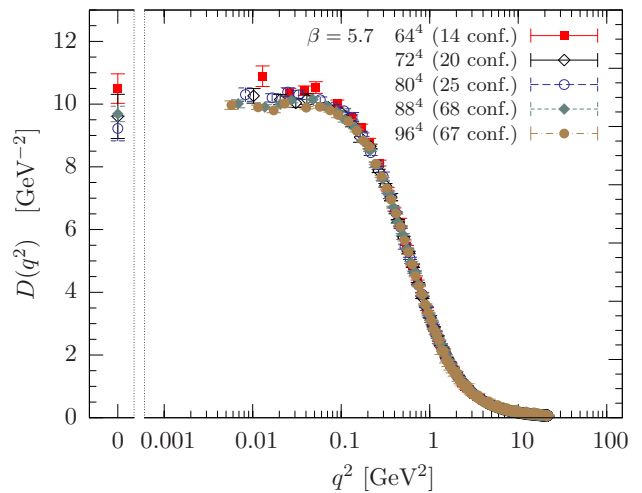}
  \caption{The bare lattice gluon propagator $D(q^2)$
    versus $q^2$ for $\beta=5.70$ and various lattice sizes. We also
    show data on $D(0)$ (left).}
  \label{fig:fig.1}
\end{figure}

\begin{figure}
  \centering
  \includegraphics[height=6.7cm]{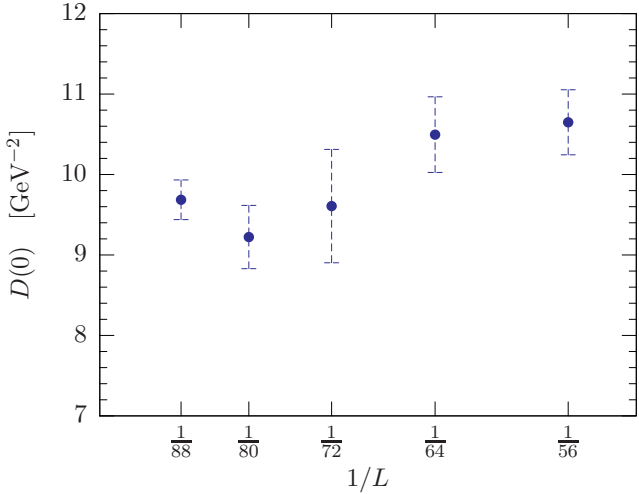}
  \caption{Zero-momentum gluon propagator $D(0)$ versus
    $1/L$.}
  \label{fig:fig.2}
\end{figure}

The gluon propagator is defined by
\begin{align} \nonumber
  D_{\mu\nu}^{ab}(q)&=\langle \widetilde{A}_{\mu}^a(k)
  \widetilde{A}_{\nu}^b(-k) \rangle\\ 
  &=\left( \delta_{\mu\nu} - \frac{q_{\mu}q_{\nu}}{q^2} \right)
  \delta^{ab} D(q^2)\,,
  \label{eq:gluonpropagator}
\end{align}
where $\widetilde{A}(k)$ represents the Fourier transform of the
gauge potentials according to \Eq{eq:gaugepotential} with Landau-gauge
fixed links. The momentum $q$ is given by
$q_{\mu}=(2/a) \sin{(\pi k_{\mu}/L)}$ with $k_{\mu} \in (-L/2,L/2]$.
For $q \ne 0$, one gets 
\begin{equation}
  D(q^2) = \frac{1}{24} \sum_{a=1}^8
  \sum_{\mu=1}^4 D^{aa}_{\mu\mu}(q) \,, 
\end{equation}
whereas at $q = 0$ the \emph{zero-momentum propagator} $D(0)$ is 
defined as
\begin{equation}
  D(0) = \frac{1}{32} \sum_{a=1}^8 \sum_{\mu=1}^4 D^{aa}_{\mu\mu}(q=0) \; .
\label{eq:zeromom_gluonpropagator}
\end{equation}
Our data for the gluon propagator obtained for various lattice sizes is 
presented\footnote{In this letter the gluon and
  ghost propagator
  data has not been renormalised in contrast to our former studies, in
  particular in \cite{Bogolubsky:2007ud}.} in \Fig{fig:fig.1}.
One can clearly see a flattening of the gluon propagator as a function
of $q^2$ for small momenta. Note also the weak volume dependence of
the results. To illustrate the latter, in \Fig{fig:fig.2} we present 
also the dependence of the zero-momentum propagator $D(0)$ (acc.\ to
\Eq{eq:zeromom_gluonpropagator}) on the inverse lattice size $1/L$. 
From this point of view it is natural to conclude that in the infinite 
volume limit the gluon propagator will approach some constant 
value as $q^2 \to 0$, i.e., the gluon dressing function 
$Z(q^2) = q^2 D(q^2)$ seems to decrease linearly with $q^2$. The 
observed IR behaviour related to the zero-momentum modes
of the lattice gauge field $A_{\mu}(x)$ can be associated with a 
massive gluon.  

Our $SU(3)$ IR gluon plateau results are in close agreement with
analogous $SU(2)$ results found recently on huge $4D$ symmetric
lattices \cite{Cucchieri:2007md,Sternbeck:2007ug}. 

\section{Ghost propagator}
\label{ghost}

The Landau-gauge ghost propagator is defined by
\begin{align} \nonumber
      G^{ab}(q) &= a^2 \sum_{x,y} \left\langle {\rm e}^{-2 \pi ik
          \cdot (x-y)/L} [M^{-1}]^{ab}_{xy} \right\rangle\\
               & = \delta^{ab}G(q^2)\,,
\end{align}
where $M$ denotes the lattice Faddeev-Popov operator, being
the Hessian of the gauge functional (\ref{eq:gaugefunctional}) 
with respect to $g_x$, in the background of the gauge-fixed links $U_{x\mu}$ 
\begin{equation}
  M^{ab}_{xy}  =  \sum_{\mu} \left[A^{ab}_{x,\mu}\,\delta_{x,y}
  - B^{ab}_{x,\mu}\,\delta_{x+\hat{\mu},y}
  - C^{ab}_{x,\mu}\,\delta_{x-\hat{\mu},y}\right]
\end{equation}
with
\begin{align*}
 A^{ab}_{x,\mu} &= \phantom{2\cdot\ } \Re\Tr\left[
 \{T^a,T^b\}(U_{x,\mu}+U_{x-\hat{\mu},\mu}) \right]\,,\\
 B^{ab}_{x,\mu} &= 2\cdot\Re\Tr\left[ T^bT^a\, U_{x,\mu}\right]\,,\\
 C^{ab}_{x,\mu} &= 2\cdot\Re\Tr\left[ T^aT^b\, U_{x-\hat{\mu},\mu}\right]\,.
\end{align*}
$T^a$, $a=1,\ldots,8$ are the (Hermitian) generators of the
$\mathfrak{su}(3)$ Lie algebra satisfying $\Tr\,[T^aT^b]=\delta^{ab}/2$.

\begin{figure}
  \centering
  \includegraphics[width=0.95\linewidth]{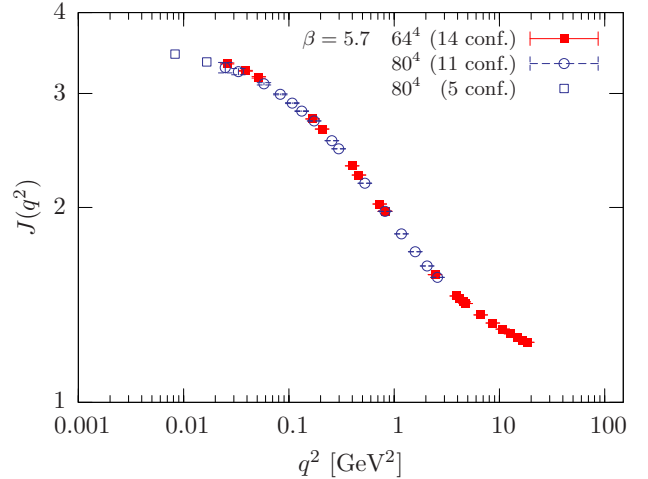}
  \caption{Bare ghost dressing function $J(q^2)$ versus $q^2$ for $L=64,80$
    at $\beta=5.70$. Errors are not shown at the two lowest
    $q^2$ (squares).}  
\label{fig:fig.3}
\end{figure}

To invert $M$ we use the conjugate gradient (CG) algorithm 
with plane-wave sources $\vec{\psi}_c$ with 
colour and position components \mbox{$\psi^a_c(x) = \delta^{ac}
\exp (2\pi i\,k\!\cdot\! x/L)$}. In fact, we apply a pre-conditioned
CG algorithm (PCG) to solve $M^{ab}_{xy}\phi^{b}(y)=\psi^a_c(x)$ where as 
pre-conditioning matrix we use the inverse Laplacian operator $\Delta^{-1}$ 
with diagonal colour substructure (for details see
\cite{Sternbeck:2005tk,Sternbeck:2006rd}).

For the large lattice sizes as considered here, we are confident 
that finite-volume distortions for all lattice momenta besides the two 
minimal ones do not change considerably with increasing $L$ (see
\Fig{fig:fig.3} and \cite{Bogolubsky:2007ud} for details). 
In this figure the ghost dressing function $J(q^2)=q^2 G(q^2)$
is presented in a log-log scale. We do not see any power-like singular 
behaviour in the limit $q^2\to 0$. Instead, we have a good indication 
that $J(q^2)$ reaches a plateau just as the \emph{decoupling solution}
of DS and FRG equations does (see also
\cite{Cucchieri:2007rg,Bornyakov:2008yx}).

\section{Running coupling}
\label{coupling}

Finally, let us present the running coupling defined as the 
renormalization group (RG) invariant product
\begin{equation}
  \label{eq:runcoupl}
  \alpha_s(q^2)= \frac{g_0^2}{4\pi} Z(q^2) J^2(q^2)\,,
\end{equation}
of the Landau-gauge gluon and ghost dressing functions. This
definition is based on the ghost-gluon vertex in a
momentum-subtraction scheme with the vertex renormalisation constant
(in Landau gauge) set to one. This is
possible \cite{Maltman:2008de}, since the vertex is known to be regular
\cite{Taylor:1971ff} (see also the lattice
studies~\cite{Cucchieri:2004sq,Ilgenfritz:2006gp}). Note that
the relation of $\alpha_s$ in this scheme to the running coupling in the
$\overline{\mathrm{MS}}$ scheme is known to four loops and it can provide a
valuable alternative to the $\overline{\mathrm{MS}}$ coupling in
phenomenological applications \cite{Maltman:2008de}.  

\begin{figure}
  \centering
  \includegraphics[width=0.95\linewidth]{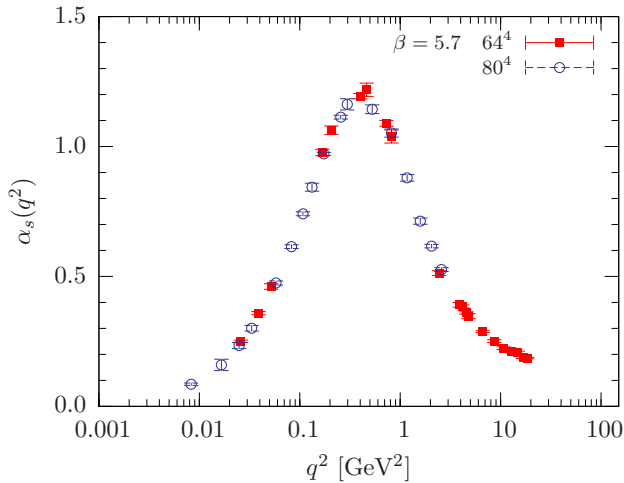} 
  \caption{Running coupling $\alpha_s(q^2)$ versus $q^2$ for lattice sizes 
    $64^4$ and $80^4$ at $\beta=5.70$.}
 \label{fig:fig.4}
\end{figure}

Beyond perturbation theory, the behaviour of $\alpha_s$ differs at
low scales for the scaling and decoupling solutions. Based on our
propagator data we can calculate $\alpha_s$ for intermediate
and lower scales, and it clearly shows a decrease towards $q^2
\to 0$ (see \Fig{fig:fig.4}). This is again consistent with the 
\emph{decoupling} DS and FRG solutions.

\section{Conclusions}
\label{conclusions}

The progress achieved on the lattice during the last two 
years in studying the IR limit of gluodynamics and checking the 
well-known scenarios of confinement in terms of Landau-gauge 
Green's functions leads us to the following conclusions.
Within the standard lattice approach as described above only the 
decoupling-type solution of DS and FRG equations seems to survive.
Since for this solution the gluon propagator tends to a non-zero
IR value, it corresponds to a massive gluon. It has been argued that
this behaviour contradicts global BRST invariance
\cite{Fischer:2008uz}.

But the lattice approach as discussed here has a few weak points.
The choice of the gauge potentials $A_{\mu}(x)$ and correspondingly 
of the gauge functional $F_U(g)$ is in no way unique. 
As long as we are reaching the infrared limit by employing rather
large lattice couplings the continuum limit is not under control. 
Moreover, we have used standard periodic boundary conditions which
certainly have an impact on the IR limit. The fact that under these
conditions the gluon propagator does not tend to zero is related to
the behaviour of the zero-momentum modes, which do not become
sufficiently suppressed as the lattice size increases. 
Changing the definition of lattice Landau gauge, and correspondingly
the lattice definitions of $A_{\mu}(x)$ and~$M$, modifying the boundary
conditions and further improving the gauge-fixing procedure, e.g., by
taking $Z(3)$ flips as mentioned in \Sec{intro} into account, may
essentially suppress the zero-momentum modes and correspondingly change
the behaviour of both the gluon and ghost propagators. Therefore, a
final conclusion still cannot be drawn.

Note that one of us (A.S.) has recently carried out a lattice
computation in the strong-coupling limit. For the gluon and the ghost
propagator at larger $a^2q^2$ it was possible to extract
the right exponents as expected for the scaling
solution~\cite{Sternbeck:2008mv}. At asymptotically small momenta,
however, results were shown to depend strongly on the lattice 
definitions of $A_{\mu}(x)$ and~$M$. This corresponds to observations in
studies of DS and FRG equations, namely that for decoupling-like solutions
any IR-asymptotic values of the gluon propagator and the ghost
dressing function can be considered as boundary conditions
\cite{Fischer:2008uz}.  

It has been argued in~\cite{vonSmekal:2008es} that a BRST-invariant 
gauge-fixing prescription is possible on the lattice.
It remains to be seen, whether the preferred scaling behaviour 
of Landau-gauge gluon and ghost propagators can be achieved 
consistently also in lattice Yang-Mills theories.

\section*{ACKNOWLEDGEMENTS}

The simulations were done on the parallel processor system MVS-15000VM 
at the Joint Supercomputer Centre (JSCC) in Moscow and on the IBM 
pSeries 690 at HLRN. This work was supported by joint grants DFG 436 RUS
113/866/0 and RFBR 06-02-04014. Part of this work is supported by
DFG under contract FOR 465 Mu 932/2-4, and by the Australian
Research Council. The authors wish to express their gratitude to
V.~Bornyakov, M.~Chernodub, A.~Dorokhov, C.~Fischer, V.~Mitrjushkin,
J.~Pawlowski, M.~Polikarpov, and L.~von Smekal for useful
discussions.



\end{document}